\documentclass[12pt,a4paper,oneside]{amsart}

\usepackage{amssymb,amsmath}
\usepackage{graphicx,psfrag}

\begin{document}

\newtheorem{thm}{Theorem}[section]
\newtheorem{definition}{Definition}[section]
\newtheorem{cor}{Corollary}[section]

\newcommand{\vx}{\bar x}
\newcommand{\vy}{\bar y}
\newcommand{\vz}{\bar z}
\newcommand{\vw}{\bar w}
\newcommand{\then}{\rightarrow}
\newcommand{\oszt}{\slash}
\newcommand{\gyok}{\sqrt{\phantom{n}}}
\newcommand{\de}{\stackrel{d}{=}}

\newcommand{\defiff}{\ \ \stackrel{d}{\Leftrightarrow}\ \ }

\newcommand{\IOb}{\ensuremath{\mathsf{IOb}}} 
\newcommand{\IB}{\ensuremath{\mathsf{IB}}} 
\newcommand{\Ob}{\ensuremath{\mathsf{Ob}}} 
\newcommand{\B}{\ensuremath{\mathit{B}}} 
\newcommand{\Ph}{\ensuremath{\mathsf{Ph}}} 
\newcommand{\F}{\ensuremath{\mathsf{F}}} 
\newcommand{\Q}{\ensuremath{\mathit{Q}}} 
\newcommand{\W}{\ensuremath{\mathsf{W}}} 
\newcommand{\SpecRel}{\ensuremath{\mathsf{SpecRel}}} 
\newcommand{\AccRel}{\ensuremath{\mathsf{AccRel}}} 
\newcommand{\GenRel}{\ensuremath{\mathsf{GenRel}}} 
\newcommand{\ev}{\ensuremath{\mathsf{ev}}} 
\newcommand{\cl}{\ensuremath{\mathsf{cl}}} 
\newcommand{\Apr}{\ensuremath{\mathsf{Apr}}} 
\newcommand{\Id}{\ensuremath{\mathsf{Id}}} 
\newcommand{\Dom}{\ensuremath{\mathsf{Dom}}} 
\newcommand{\M}{\ensuremath{\mathfrak{M}}} 
\newcommand{\ax}[1]{\ensuremath{\mathsf{#1}}} 
\newcommand{\AxEv}{\ensuremath{\mathsf{AxEv}}}
\newcommand{\wl}{\ensuremath{\mathsf{wl}}}
\newcommand{\w}{\ensuremath{\mathsf{w}}}
\newcommand{\AxDif}{\ensuremath{\mathsf{AxDf}}}
\newcommand{\AxCmv}{\ensuremath{\mathsf{AxCm}}}
\newcommand{\AxCont}{\ensuremath{\mathsf{Cont}}}
\newcommand{\AxEOb}{\ensuremath{\mathsf{AxEOb}}}
\newcommand{\AxField}{\ensuremath{\mathsf{AxFd}}}
\newcommand{\AxSelf}{\ensuremath{\mathsf{AxSf}}}
\newcommand{\AxPh}{\ensuremath{\mathsf{AxPh}}}
\newcommand{\AxSym}{\ensuremath{\mathsf{AxSym}}}
\newcommand{\AxSimd}{\ensuremath{\mathsf{AxSm}}}

\author{Hajnal Andr\'eka, Judit X.\ Madar\'asz, Istv\'an N\'emeti and
  Gergely Sz\'ekely} \thanks{This research is supported by the
  Hungarian Scientific Research Fund for basic research grants
  No.~T73601 and T81188, as well as by a Bolyai grant for J.~X.~Madar\'asz}
\title[A logic road from special to general relativity]{A logic road
  from special relativity to general relativity}

\maketitle

\begin{abstract}
We present a streamlined axiom system of special relativity in
first-order logic.  From this axiom system we ``derive'' an axiom
system of general relativity in two natural steps. We will also see
how the axioms of special relativity transform into those of general
relativity. This way we hope to make general relativity more
accessible for the non-specialist.
\end{abstract}

\section*{Introduction}

In axiomatizing physical theories, we follow in the footsteps of
many great predecessors. Logical axiomatization of physics,
especially that of relativity theory, is not at all a new idea. It
goes back to such leading mathematicians and philosophers as
Hilbert, G{\" o}del, Tarski, Reichenbach, Carnap, Suppes and
Friedman. It also has an extensive literature, see, e.g., the
references of \cite{pezsgo}, \cite{AMNsamples}. Our aims go beyond
these approaches, because we not only axiomatize relativity
theories, but also analyze their logical and conceptual structures.

There are many examples showing the benefits of using the axiomatic
method in the foundations of mathematics. The success story of
the axiomatic method in the foundations of mathematics suggests that it
is worth to apply this method in the foundations of spacetime
theories.

For good reasons, foundations of mathematics was carried through
strictly within first-order logic (FOL).  For the same reasons,
foundations of spacetime theories are best developed within FOL. For
example, in any foundational work it is essential to avoid tacit
assumptions, and one acknowledged feature of using FOL is that it
helps to eliminate tacit assumptions. That is only one of the many
reasons why we work within FOL. For further reasons, see \cite[\S
Why FOL?]{pezsgo}, \cite[\S 11]{SzDis}.

In physics, the same way as in mathematics, we do not address the
question whether the axioms are true or not, we just postulate them.
The reason for this in mathematics is that we want to give a tool
that is usable in all applications where the axioms are true.
However, in physics the statements of the theories are closely
related to the real physical world, thus the application area is
fixed in a way. Therefore, the role of the axioms (the role of
statements that we assume without proofs) in physics is more
fundamental than in mathematics. That is why we aim to formulate
simple, logically transparent and intuitively convincing axioms. All
the surprising or unusual predictions of a physical theory should be
provable as theorems and not assumed as axioms. For example, the
prediction ``no observer can move faster than light'' is a theorem
in our approach and not an axiom, see Theorem~\ref{noftl-thm}.

Some of the questions we study when investigating the logical
structure of relativity theories are: -- What is believed and why?
-- Which axioms are responsible for certain predictions? -- What
happens if we discard some axioms?  -- Can we change the axioms and
at what price?

First-order logic can be viewed as a fragment of natural language
with unambiguous syntax and semantics. Being a {\it fragment of
natural language} is useful in our project because one of our aims
is to make relativity theory accessible to a broad audience. {\it
Unambiguous syntax and semantics} are  important for the same
reason, because they make it possible for the reader to always know
what is stated and what is not stated by the axioms. Therefore they
can use the axioms without being familiar with all the tacit
assumptions and rules of thumb of physics (that one usually learns
via many, many years of practice).

A novelty in this paper is that we concentrate on the transition
from special relativity to general relativity, we try to keep this
transition logically transparent and illuminating for the
non-specialist. We are going to ``derive'' the axioms of general
relativity from those of special relativity in two natural steps. In
the first step we will extend our FOL axiom system of special
relativity of inertial observers to accelerated observers. This step
will provide us a FOL theory of accelerated observers, which implies
the usual predictions about them, such as the twin paradox, see
Theorem~\ref{thm-twp}.  In the second step we will eliminate the
difference between inertial and noninertial observers on the level
of axioms. By these two natural steps, we will get a FOL
axiomatization of the spacetimes of {\it general relativity}
suitable for further study. All these three theories (special
relativity, theory of accelerated observers, general relativity)
will be formulated in the same streamlined FOL language.

\section{The FOL language of our three theories}
\label{lang-s}

The first important decision in writing up an axiom system in FOL is
to choose the vocabulary (set of basic symbols) of our language, i.e.,
what objects and relations between them we will use as basic
concepts. We will have to stick to this language while writing up the
axioms and investigating our theory. However, later we can change the
vocabulary of our language, and we can write up a new theory (or a new
version of our theory) in this new language. Then we can investigate
the logical connections between these theories built up in different
FOL languages. This way we can investigate the role of having chosen
the particular basic concepts of our theory, see \cite[second half of
  \S 3]{WKU09}.

In this paper we will use the following two-sorted FOL language:
\begin{equation*}
\{\, \B, \IB, \Ph,\; \Q,+,\cdot,\; \W\,\},
\end{equation*}
where $\B$ (bodies\footnote{By bodies we mean anything which can move,
  e.g., test-particles, reference frames, electromagnetic waves,
  centers of mass, etc.}) and $\Q$ (quantities) are the two sorts,
$\IB$ (inertial bodies) and $\Ph$ (light signals or
photons\footnote{Here we use light signals and photons as synonyms
  because it is not important here whether we think of them as
  particles or electromagnetic waves. The only thing that matters here is that
  they are things that can move. So they are bodies in the sense of
  our FOL language.}) are one-place relation symbols of sort $\B$, $+$
and $\cdot$ are two-place function symbols of sort $\Q$, and $\W$ (the
worldview relation) is a $2+4$-place relation symbol the first two
arguments of which are of sort $\B$ and the rest are of sort $\Q$.

Atomic formulas $\IB(b)$ and $\Ph(p)$ are translated as ``\textit{$b$
  is an inertial body},'' and ``\textit{$p$ is a photon},''
respectively.  We use the worldview relation $\W$ to speak about
coordinatization by translating $\W(o,b,x,y,z,t)$ as ``\textit{body $o$
  coordinatizes body $b$ at space-time location $\langle
  x,y,z,t\rangle$},'' (i.e., at space location $\langle x,y,z\rangle$
and at instant $t$). Sometimes we use the more picturesque
expressions \textit{sees} or \textit{observes} for coordinatizes.
However, these ``seeing'' and ``observing'' have nothing to do with
visual seeing or observing they only mean associating coordinate
points to bodies.

The above, together with statements of the form $x=y$ (read as
\textit{$x$ equals $y$}) are the so-called \textit{atomic formulas} of
our FOL language, where $x$ and $y$ can be arbitrary variables of the
same sort, or terms built up from variables of sort \Q\ by using the
two-place operations $\cdot$ and $+$. The \textit{formulas} of our FOL
language are built up from these atomic formulas by using the logical
connectives \textit{not} ($\lnot$), \textit{and} ($\land$),
\textit{or} ($\lor$), \textit{implies} ($\rightarrow$),
\textit{if-and-only-if} ($\leftrightarrow$) and the quantifiers
\textit{exists} ($\exists$) and \textit{for all} ($\forall$).  For the
precise definition of the syntax and semantics of FOL, see, e.g.,
\cite[\S 1.3]{CK}, \cite[\S 2.1, \S 2.2]{End}, or
\cite[pp.39--46]{HMT}.

For example, the formula $\forall
btxzy\enskip\W(b,b,x,y,z,t)\rightarrow x=y$ abbreviates the natural
language sentence ``\textit{For all $b$, $t$, $x$, $y$, $z$ it is true
  that $b$ observes $b$ at $t$, $x$, $y$, $z$ implies that $x$ equals
  $y$},'' or in a more readable form ``\textit{If $b$ sees itself
  at $t$, $x$, $y$, $z$, then $x$ equals $y$; and this is true for all
  $b$, $t$, $x$, $y$, $z$.}''

To abbreviate formulas of FOL we often omit parentheses according to
the following convention. Quantifiers bind as long as they can, and
$\land$ binds stronger than $\rightarrow$. For example, we write
$\forall x\enskip \varphi\land\psi\rightarrow\exists y\enskip
\delta\land\eta$ instead of $\forall
x\big((\varphi\land\psi)\rightarrow\exists y(\delta\land\eta)\big)$.

We will use the letters $p$, $b$, $m$, $k$, $h$ and their
variants for variables of sort $\B$, and the letters $x$, $y$, $z$,
$t$, $v$, $w$, $c$ and their variants for variables of sort $\Q$. For
easier readability, we will use $\vx$, $\vy$, etc. for sequences of
variables.  The $i$-th element of the sequence $\vx$ is denoted
by $x_i$.

\section{An axiomatization of special relativity in FOL}
\label{ax-s}

Having fixed  our language, we now turn to formulating an axiom
system for special relativity in this language. The first axiom
states some usual properties of addition $+$ and multiplication
$\cdot$ true for real numbers.

\begin{itemize}
\item[\underline{\AxField:}]
 The quantity part $\langle \Q,+,\cdot\rangle$ is a Euclidean field, i.e.,
\item $\langle\Q,+,\cdot\rangle$ is a field in the sense of abstract
algebra,
\item the relation $\le$ defined by $\,x\le y\defiff \exists
  z\enskip x+z^2=y\,$ is a linear ordering on $\Q$, and
\item $\forall x\enskip\exists y\enskip x=y^2\lor-x=y^2$, i.e.,
positive elements are squares.\footnote{We note that the second statement in the definition of
\AxField\ can be replaced with $x^2+y^2+z^2=0\rightarrow x=0$.}
\end{itemize}
\noindent The field-axioms (see, e.g., \cite[pp.40--41]{CK},
\cite[p.38]{Hodges}) say that $+$, $\cdot$ are associative and
commutative, they have neutral elements $0$, $1$ and inverses $-$,
$\oszt$ respectively, with the exception that $0$ does not have an
inverse with respect to $\cdot\,$, as well as $\cdot$ is additive
with respect to $+$. We will use $0$, $1$, $-$, $\oszt$, $\gyok$ as
derived (i.e., defined) operation symbols. As usual, $x^2$ denotes
$x\cdot x$.

\AxField\ is a ``mathematical" axiom in spirit. However, it has
physical (even empirical) relevance. Its physical relevance is that
we can add and multiply the outcomes of our measurements and some
basic rules apply to these operations. Physicists use all properties
of the real numbers tacitly, without stating explicitly which
property is assumed and why. The two properties of real numbers
which are the most difficult to defend are the Archimedean property,
see \cite{Rosinger08}, \cite[\S  3.1]{Rosinger09}, and the supremum
property,\footnote{The supremum
  property (i.e., every nonempty and bounded subset of the real
  numbers has a least upper bound) implies the Archimedean property. So
  if we want to get ourselves free from the Archimedean property, we have
  to leave this property, too.} see the remark after the introduction of \ax{Cont} on
p.\pageref{p-cont}.

In special relativity, we will not need more properties of the real
numbers than stated in \AxField. In the next two sections, we will use
more properties of the real numbers in our next two theories for
relativity, but we will state exactly and explicitly how much we will use.

Euclidean fields got their names after their role in Tarski's FOL
axiomatization of Euclidean geometry \cite{TarskiElge}. By
\AxField\ we can reason about the Euclidean structure of a coordinate
system the usual way, we can introduce Euclidean distance, talk about
straight lines, etc.  In particular, we will use the following
notation for $\vx,\vy\in\Q^n$ if $n\ge 1$:
\begin{equation*}
|\vx|=\sqrt{x_1^2+\dots+x_n^2},\quad\text{ and }\quad
  \vx-\vy\de\langle x_1-y_1,\dots,x_n-y_n\rangle.
\end{equation*}

The rest of our axioms speak about the worldviews of inertial
observers. We have not introduced the concept of observers as a
basic one because it can be defined as follows: an {\it observer} is
nothing else than a body who ``observes" (coordinatizes) some other
bodies somewhere, this property can be captured by the following
formula of our language:
\begin{equation*}
\Ob(m) \defiff  \exists b\vx\enskip \W(m,b,\vx);
\end{equation*}
and {\it inertial observers} can be defined as inertial bodies which
are observers, formally:
\begin{equation*}
\IOb(m) \defiff \IB(m)\land \Ob(m).
\end{equation*}

\noindent
We will also use the following two notations:
\begin{equation*}
\vx_s\de \langle x_1,x_2, x_3\rangle \quad \text{ and }\quad {x_t}\de x_4
\end{equation*}
for the \textit{space component} and the
\textit{time component} of $\vx\in\Q^4$,
respectively.

Our next axiom is the key axiom of $\SpecRel$, it has an immediate
physical meaning. This axiom is the outcome of the Michelson-Morley
experiment. It has been continuously tested ever since then.
Nowadays it is tested by GPS technology.

\begin{itemize}
\item[\underline{\AxPh:}] For any inertial observer, the speed of light is the same
  everywhere and in every direction, and it is finite. Furthermore, it
  is possible to send out a light signal in any direction. Formally:
\begin{multline*}
\forall m\enskip\exists c_m\enskip\forall \vx\vy \enskip \IOb(m)\rightarrow\\
\big(\exists p \enskip \Ph(p)\land \W(m,p,\vx)\land \W(m,p,\vy)\big)
\leftrightarrow |\vy_s-\vx_s|= c_m\cdot|y_t-x_t|.
\end{multline*}
\end{itemize}

Let us note here that \ax{AxPh} does not require that the speed of
light is the same for every inertial observer or that it is nonzero.
It requires only that the speed of light according to a fixed
inertial observer is a quantity which does not depend on the
direction or the location.

Our next axiom connects the worldviews of different inertial
observers by saying that all observers observe the same ``external"
reality (the same set of events). Intuitively, by the event
occurring for $m$ at $\vx$, we mean the set of bodies $m$ observes
at $\vx$. Formally:
\begin{equation*}
\ev_m(\vx)\de\{ b : \W(m,b,\vx)\}.
\end{equation*}

\begin{itemize}
\item[\underline{\AxEv:}]
All inertial observers coordinatize the same set of events, i.e.,
\begin{equation*}
\forall mk\enskip \IOb(m)\land\IOb(k)\enskip\rightarrow\enskip
\forall \vx\enskip \exists \vy\enskip \forall b\enskip
\W(m,b,\vx)\leftrightarrow\W(k,b,\vy).
\end{equation*}
\end{itemize}
Hereafter, we will use $\ev_m(\vx)=\ev_k(\vy)$ to abbreviate the
subformula $\forall b\enskip \W(m,b,\vx)\leftrightarrow\W(k,b,\vy)$
of \ax{AxEv}.

Our two remaining axioms are simplifying ones. We could leave them
out without losing the essence of our theory, only the
formalizations of the theorems would become much more complicated.

\begin{itemize}
\item[\underline{\AxSelf:}]
Any inertial observer sees himself as standing still at the origin:
\begin{equation*}
\forall m\enskip\IOb(m)\rightarrow\enskip \big(\forall \vx\enskip
\W(m,m,\vx) \leftrightarrow x_1=0\land x_2=0\land x_3=0\big).
\end{equation*}
\end{itemize}

Our last axiom is a symmetry axiom saying that all observers use the
same units of measurement.

\begin{itemize}
\item[\underline{\AxSimd:}]
Any two inertial observers agree as to the spatial distance between
two events if these two events are simultaneous for both of them;
furthermore, the speed of light is 1 for all observers:
\begin{multline*}
\forall mk\enskip\IOb(m)\land\IOb(k)\rightarrow \forall
\vx\vy\vx'\vy' \enskip x_t=y_t\land x'_t=y'_t\land\\
\ev_m(\vx)=\ev_k(\vx')\land \ev_m(\vy)=\ev_k(\vy')\rightarrow
|\vx_s-\vy_s|=|\vx'_s-\vy'_s|, \text{ and }
\end{multline*}
\begin{multline*}
\forall
m\enskip\IOb(m)\rightarrow\exists
p\enskip\Ph(p)\land\W(m,p,0,0,0,0)\land\W(m,p,1,0,0,1).
\end{multline*}
\end{itemize}

We introduce an axiom system for special relativity as the collection of these five axioms:
\begin{equation*}
\SpecRel \de \{ \AxField, \AxPh, \AxEv, \AxSelf, \AxSimd\}.
\end{equation*}

The so-called worldline of body $b$ according to observer $m$ is
defined as follows:
\begin{equation*}
\wl_m(b)\de\{ \vx: \W(m,b,\vx)\}.
\end{equation*}

To abbreviate formulas, we will use bounded quantifiers in the
following way: $\exists x\; \varphi(x)\land \psi$ and $\forall x\;
\varphi(x)\rightarrow \psi$ are abbreviated to $\exists
x\in\varphi\enskip \psi$ and $\forall x\in\varphi\enskip \psi$,
respectively. For example, $\forall \vx,\vy\in\wl_m(b) \enskip \psi$
abbreviates $\forall\vx\vy\enskip \W(m,b,\vx)\land\W(m,b,\vy) \then
\psi$.\footnote{Both $b\in\ev_m(\vx)$ and $\vx\in\wl_m(b)$ represent
  the same atomic formula of our FOL language, namely: $\W(m,b,\vx)$.}

In an axiom system, the axioms are the ``price" we pay, and the
theorems are the ``goods" we get for them. Therefore, we strived for
putting only simple, transparent, easy-to-believe statements in our
axiom system. We want to get all the hard-to-believe predictions as
theorems. For example, now we are going to prove from $\SpecRel$
that it is impossible for inertial observers to move faster than
light relative to each other. This theorem is a generic example for
a ``fancy theorem'' following from ``plain axioms.''

\begin{thm}{\rm (no faster than light inertial observers)} \label{noftl-thm}
\begin{multline*}
\SpecRel\vdash \forall mk\enskip \forall
\vx,\vy\in\wl_m(k)\quad\\ \vx\neq\vy\land\IOb(m)\land\IOb(k) \then
|\vy_s-\vx_s|<|y_t-x_t|.
\end{multline*}
\end{thm}

Intuitively, no observer can travel faster than light relative to
another. Let us see the axioms in action: now we will prove
Theorem~\ref{noftl-thm} paying a close attention to the axioms used
in each step.
\begin{proof}
Let $m$ and $k$ be inertial observers and let $\vx,\vy\in\wl_m(k)$
such that $\vx\neq\vy$.  By \ax{AxFd}, $\le$ is a total order, so
there are three possibilities only: $|\vy_s-\vx_s|<|y_t-x_t|$,
$|\vy_s-\vx_s|>|y_t-x_t|$ or $|\vy_s-\vx_s|=|y_t-x_t|$.  We will
prove $|\vy_s-\vx_s|<|y_t-x_t|$ by excluding the other two
possibilities.

\begin{figure}
\psfrag{m}[l][l]{$m$} \psfrag{k}[l][l]{$k$} \psfrag{p}[br][br]{$p$}
\psfrag{p1}[tl][tl]{$p_1$} \psfrag{p2}[l][l]{$p_2$}
\psfrag{p3}[l][l]{$p_3$} \psfrag{x}[tl][tl]{$\vx$}
\psfrag{y}[br][br]{$\vy$} \psfrag{yt}[r][r]{$y_t$}
\psfrag{z}[b][b]{$\vz$} \psfrag{w}[tl][tl]{$\vw$}
\psfrag{xp}[tl][tl]{$\vx'$} \psfrag{yp}[bl][bl]{$\vy'$}
\psfrag{wp}[t][t]{$\vw'$} \psfrag{zp}[l][l]{$\vz'$}
\psfrag{zs}[l][l]{$\vz_s$} \psfrag{ys}[tl][tl]{$\vy_s$}

\includegraphics[keepaspectratio, width=\textwidth]{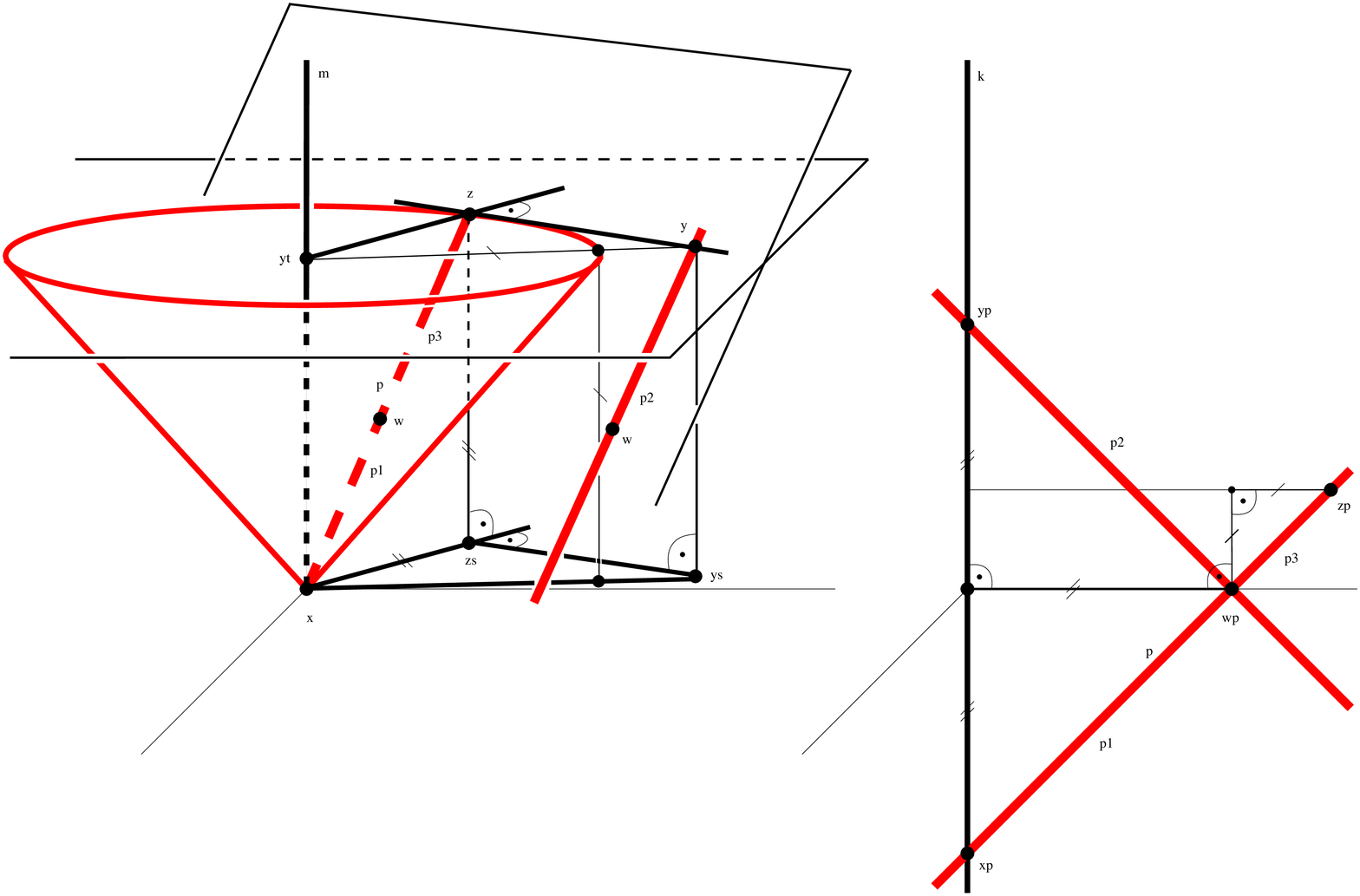}
\caption{\label{fig-noftl} Illustration for the proof of
  Theorem~\ref{noftl-thm}}
\end{figure}

Let us first prove that $|\vy_s-\vx_s|>|y_t-x_t|$ cannot hold.  Figure
\ref{fig-noftl} illustrates this proof.\footnote{To simplify the
  figure, we have drawn $\vx$ to the origin. This is not used in the
  proof but it can be assumed without losing generality.}  So, let us
assume that $|\vy_s-\vx_s|>|y_t-x_t|$, we will derive a
contradiction. By \ax{AxFd}, there is a coordinate point $\vz$ such
that $|\vz_s-\vx_s|=|z_t-x_t|\neq 0$, $z_t=y_t$ and $\vz_s-\vx_s$ is
orthogonal to $\vz_s-\vy_s$ if $x_t\neq y_t$, and
$|\vz_s-\vx_s|=|z_t-x_t|\neq 0$ and $\vz_s-\vx_s$ is orthogonal to
$\vy_s-\vx_s$ if $x_t=y_t$ (here we used that
$|\vy_s-\vx_s|>|y_t-x_t|$). Any choice of such a $\vz$ implies that
any line of slope $1$ in the plane $\vx\vy\vz$ is parallel to the line
$\vx\vz$ (because the plane $\vx\vy\vz$ is tangent to the light cone
through $\vz$).  To choose one concrete $\vz$ from the many,
let \begin{equation*}\vw_s \de \frac{\vy_s-\vx_s}{|\vy_s-\vx_s|},\quad
  \vw_s^{\perp}\de\frac{\langle
    y_2-x_2,x_1-y_1,0\rangle}{\sqrt{(y_2-x_2)^2+(x_1-y_1)^2}}.
\end{equation*}
Then, if $x_t=y_t$, let 
 \begin{equation*}
\vz_s\de|\vy_s-\vx_s|\cdot \vw_s^{\perp}
+\vx_s, \quad z_t\de|\vy_s-\vx_s|+x_t,
\end{equation*}
and, if $x_t\neq y_t$, let
\begin{equation*}
\vz_s\de\frac{|y_t-x_t|^2}{|\vy_s-\vx_s|}\cdot \vw_s +
\frac{|y_t-x_t|\cdot\sqrt{|\vy_s-\vx_s|^2-|y_t-x_t|^2}}{|\vy_s-\vx_s|}\cdot\vw_s^{\perp},\quad
z_t\de y_t.
\end{equation*}
See Figure~\ref{fig-z}.  It is easy to see that this $\vz$ has the
required properties.

Then by \ax{AxPh} and \ax{AxSm}, there is a
photon $p$ such that $p\in \ev_m(\vx)\cap \ev_m(\vz)$ since
$|\vz_s-\vx_s|=|z_t-x_t|$. Also by \ax{AxPh}, \ax{AxSm} and
\ax{AxFd}, there are photons showing that inertial observers see
distinct events in distinct points, so $\ev_m(\vx)$, $\ev_m(\vy)$
and $\ev_m(\vz)$ are distinct.  By \ax{AxEv}, there are coordinate
points $\vx'$, $\vy'$ and $\vz'$ such that $\ev_m(\vx)=\ev_k(\vx')$,
$\ev_m(\vy)=\ev_k(\vy')$ and $\ev_m(\vz)=\ev_k(\vz')$.  We have that
$\vx'$, $\vy'$ and $\vz'$ are distinct since $\ev_m(\vx')$,
$\ev_m(\vy')$ and $\ev_m(\vz')$ are so. By \ax{AxSf},
$\vx'_s=\vy'_s=\langle0,0,0\rangle$. By \ax{AxPh} and \ax{AxSm},
$|\vz'_s-\vx'_s|=|z'_t-x'_t|$. By \ax{AxFd}, there is a coordinate
point $\vw'$ on the line $\vx'\vz'$ such that
$|\vx'_s-\vw'_s|=|x'_t-w'_t|$, $|\vy'_s-\vw'_s|=|y'_t-w'_t|$ and
$|\vz'_s-\vw'_s|=|z'_t-w'_t|$.  By \ax{AxPh} and \ax{AxSm}, there
are photons $p_1$, $p_2$ and $p_3$ such that $\vx',\vw'\in
\wl_k(p_1)$, $\vy',\vw'\in \wl_k(p_2)$ and $\vw',\vz'\in
\wl_k(p_3)$. By \ax{AxEv}, there should be a coordinate point $\vw$
such that $\ev_m(\vw)=\ev_k(\vw')$. By \ax{AxPh} and \ax{AxSm}, this
$\vw$ should be on the line $\vx\vz$, since by \ax{AxFd} there is no
nondegenerate triangle whose sides are of slope $1$ (this fact can
be shown by proving that if we project a triangle of this kind
vertically, we get another triangle whose one side is the sum of the
other two). Specially, $\vw$ should be in the plane $\vx\vy\vz$. By
\ax{AxPh} and \ax{AxSm}, this $\vw$ should also be on a line of
slope $1$ through $\vy$. Since $\vw$ is in the plane $\vx\vy\vz$,
line $\vw\vy$ has to be parallel to the line $\vx\vz$. However,
distinct parallel lines do not intersect. Thus this $\vw$ cannot
exist. That contradicts \ax{AxEv}, so $|\vy_s-\vx_s|>|y_t-x_t|$
cannot hold.

\begin{figure}
\psfrag{xs}[tl][tl]{$\vx_s$}
\psfrag{ws}[tl][tl]{$\vw_s$}
\psfrag{wsm}[br][br]{$\vw_s^{\perp}$}
\psfrag{xp}[tl][tl]{$\vx'$}
\psfrag{yp}[bl][bl]{$\vy'$}
\psfrag{wp}[t][t]{$\vw'$}
\psfrag{zp}[l][l]{$\vz'$}
\psfrag{zs}[l][l]{$\vz_s$}
\psfrag{ys}[tl][tl]{$\vy_s$}
\psfrag{xt-yt}[br][br]{$|y_t-x_t|$}
\psfrag{xtxyt}[l][l]{$x_t\neq y_t$}
\psfrag{xt=yt}[l][l]{$x_t=y_t$}
\includegraphics[keepaspectratio, width=\textwidth]{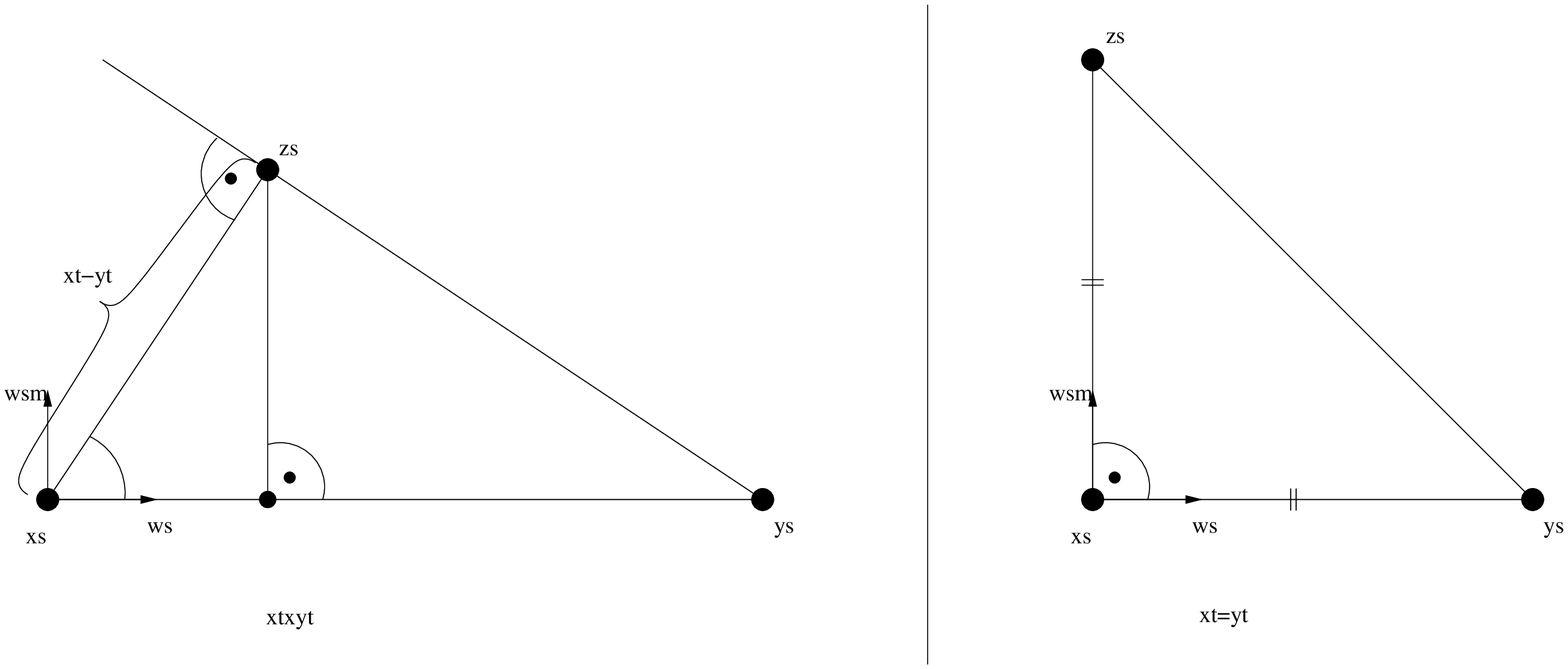}
\caption{\label{fig-z} Illustration for the proof of
  Theorem~\ref{noftl-thm}}
\end{figure}

Let us now prove that $|\vy_s-\vx_s|=|y_t-x_t|$ cannot hold, either.
If $|\vy_s-\vx_s|=|y_t-x_t|$, then by \ax{AxPh} and \ax{AxSm}, there
is a photon $p$, such that $p\in \ev_m(\vx)\cap\ev_m(\vy)$. By
\ax{AxPh}, \ax{AxSm} and \ax{AxFd}, $\ev_m(\vx)\neq\ev_m(\vy)$ since
$\vx\neq\vy$.  So by $\ax{AxEv}$, there are distinct coordinate
points $\vx'$ and $\vy'$ such that $\ev_m(\vx)=\ev_k(\vx')$ and
$\ev_m(\vy)=\ev_k(\vy')$. By \ax{AxSf},
$\vx'_s=\vy'_s=\langle0,0,0\rangle$. So $|\vy'_s-\vx'_s|=0$. By
\ax{AxPh} and \ax{AxSm}, $|\vy'_s-\vx'_s|=|y'_t-x'_t|$.  Hence
$y'_t=x_t'$, too. Thus $\vy'=\vx'$. That contradicts the fact that
$\vx'$ and $\vy'$ are distinct.

The only remaining possibility is that $|\vy_s-\vx_s|<|y_t-x_t|$
which was to be proved.
\end{proof}

It is easy to see that \ax{AxSm} was not fully used in the proof
above. We only used its second part, i.e., that the speed of light
(the $c_m$ in \ax{AxPh}) is $1$. Even the use of this part of
\ax{AxSm} was not essential. By slightly changing the proof, we
could get basically the same result using only that $c_m\neq0$, but
then we also need to mention the speed $c_m$ of light (according to
$m$) explicitly in the formalization of the theorem.

In relativity theory we are often interested in comparing the
worldviews of two different observers. To do so, we introduce the
worldview transformation between observers $m$ and $k$  (in symbols, $\w_{mk}$) as the
following binary relation:
\begin{equation*}
\w_{mk}(\vx,\vy)\defiff \ev_m(\bar
x)=\ev_k(\vy)\neq\emptyset.
\end{equation*}

By the following theorem, the worldview transformations between
inertial observers in the models of \ax{SpecRel} are not only binary
relations but very special transformations. For the definition of a
Poincar{\'e} transformation (which is a Lorentz transformation
composed with a translation) we refer to \cite[p.110]{dinverno} or
to \cite[pp.66--69]{MTW}. For the proof of the next theorem, see
\cite[Thm.11.10, p.640]{logst} or \cite[Thm.3.2.2, p.22]{SzDis}.

\begin{thm}\label{thm-poi}
\begin{equation*}
\SpecRel \vdash \forall m,k \enskip \IOb(m)\land \IOb(k) \then \w_{mk}
\text{ \rm is a Poincar{\'e} transformation.}
\end{equation*}
\end{thm}

Every Poincar{\'e} transformation is an affine\footnote{Let us recall
  that an affine transformation is the composition of a linear mapping
  and a translation.} one, specially it takes lines to lines. So by
\ax{AxSf}, Theorem~\ref{thm-poi} implies that $\wl_m(k)$ is a line
for any inertial observers $m$ and $k$.  Thus Theorem~\ref{thm-poi}
implies Theorem~\ref{noftl-thm} since a Poincar{\'e} transformation
cannot take a line of slope less than $1$ (slower than light) to a
line of slope more than $1$ (faster than light). In the proof of
Thm.\ref{thm-poi} the symmetry axiom, \ax{AxSm} has to be fully
used.\footnote{Without \ax{AxSm} there are other possible worldview
  transformations, such as dilations (i.e., rescaling of the units of
  measurement). For a complete characterization of the possible
  worldview transformations without \ax{AxSm}, see, e.g.,
  \cite[Thm.1.2(i)]{AMNsamples}.} Therefore, using Theorem~\ref{thm-poi} to prove that ``no
inertial observer can move faster than light'' does not show the
roles of the particular axioms in the proof, e.g., it does not
reveal that the first part of \ax{AxSm} plays no role in proving the
no faster than light theorem.

By Theorem \ref{thm-poi}, \ax{SpecRel} implies the paradigmatic
effects of special relativity, i.e., ``moving clocks slow down,''
``moving meter-rods shrink'' and ``moving pairs of clocks get out of
synchronism.''  However, we prefer proving these effects directly from
the axioms, see, e.g., \cite[Thm.11.6, p.631]{logst}, \cite{AMNSzCN}.

\section{Axioms for accelerated observers in FOL}
\label{acc-s}

In $\SpecRel$ we restricted our attention to inertial observers. It is
a natural idea to generalize the theory by including accelerated
observers as well. We will refer to such a generalized theory as a
theory of accelerated observers. It is explained in the classic
textbook \cite[pp.163--165]{MTW} that the study of accelerated
observers can be regarded as a natural first step (from special
relativity) towards general relativity. The theory of accelerated
observers can also be used to explain how the relativistic
paradigmatic effects of special relativity develop.

The most important axiom for accelerated observers will state that
at each moment of his life-time, an accelerated observer
coordinatizes the world near him for a short while as some inertial
observer does. How can we formalize this idea? Saying that the
worldview transformation $\w_{mk}$ is the identity function in a
neighborhood would state that $m$ and $k$ totally agree in this
neighborhood. This statement would connect the worldviews too
rigidly. We want to state a somewhat looser connection between $m$
and $k$. So, instead, we will state that the identity function
approximates the worldview transformation $\w_{mk}$ at the spacetime
point in question.

Let $f,g:\Q^n\to \Q^m,\ n,m\ge 1$ be partial\footnote{Partial means
that $f$ and $g$ are not
  necessarily everywhere defined on $\Q^n$.}  mappings and
$\vx\in\Q^n$. We say that \textit{$f$ approximates $g$ at $\vx$}, in
symbols $f\sim_{\vx} g$, if
\begin{equation*}
\forall \varepsilon >0\enskip\exists \delta>0\enskip\forall
\vy\enskip |\vy-\vx|\le\delta\rightarrow |f(\vy)-g(\vy)|\le\varepsilon\cdot
|\vy-\vx|.
\end{equation*}

Let us recall that for partial functions $f$ and $g$, the formula
$|f(\vy)-g(\vy)|\le z$ is true iff both $f(\vy)$ and $g(\vy)$ are
defined and the inequality holds. Thus $f\sim_{\vx} g$ implies that
$\vx$ has a neighborhood where both $f$ and $g$ are defined; and also
it implies that $f(\vx)=g(\vx)$.

Let us note that approximation at a given point is an equivalence
relation on functions, and if two affine mappings approximate each
other, then they are equal. These can be proved from $\AxField$.

Let $\Id$ denote the identity function from $\Q^4$ to $\Q^4$, i.e.,
$\Id(\vx)=\vx$ for all $\vx\in\Q^4$.

\begin{itemize}
\item[\underline{\AxCmv:}] At each moment of his life, observer $k$ ``sees" (i.e.,
  coordinatizes) the nearby world for a short while as an inertial
  observers $m$ does, i.e., the identity map $\Id$ approximates
  $\w_{mk}$ at this moment:
\begin{equation*}
\forall k\in \Ob\enskip\forall \vx\in \wl_k(k)\enskip\exists m\in
\IOb\enskip \w_{mk}\sim_{\vx}\Id.
\end{equation*}
\end{itemize}

Let us note here that  \ax{AxCm} is true for inertial observers in a stronger form, i.e.,
$\SpecRel \vdash \forall k \enskip\IOb(k)\then \w_{kk}=\Id$.
\noindent Axiom \ax{AxCm} ties the behavior of accelerated observers
to those of inertial ones. Justification of this axiom is given by
experiments. If $\w_{mk}\sim_{\vx}\Id$, we say that $m$ and $k$
\textit{comove} at $\vx$.  If $k$ is an accelerated spaceship, we can
think of a dropped spacepod as a comoving inertial observer (comoving
at the moment of dropping). Or, if $k$ switches off his engines at
$\vx$, he will move on as a comoving inertial observer would.

Assuming \SpecRel, \ax{AxCm} implies that the worldlines of $m$ and
$k$ meet and are tangent at $\vx$ in the worldviews of all other
inertial observers. Moreover, any body $b$ present in the event
at $\vx$, $\wl_m(b)$ and $\wl_k(b)$ are tangent at $\vx$;
intuitively we could say that the whole worldviews of $m$ and $k$ are
tangent at $\vx$.

Our next two axioms ensure that the worldviews of accelerated
observers are big enough. They are generalized versions of the
corresponding axioms for inertial observers, but now postulated for
all observers.

\begin{itemize}
\item[\underline{$\AxEv^-$:}] If $m$ sees $k$ participate in an event, then $k$
cannot deny it, i.e.,
\begin{equation*}\forall m,k\in\Ob\enskip  \W(m,k,\vx)\rightarrow\exists
\vy\enskip \ev_m(\vx)=\ev_k(\vy).
\end{equation*}
\item[\underline{$\AxSelf^-$:}] The worldline of any observer is an interval
of the time-axis, in his own worldview:
\begin{multline*}
 \forall m\in\Ob\enskip\forall \vx\enskip \W(m,m,\vx)\then x_1=x_2=x_3=0
 \quad \text{ and }\\
\forall \vx,\vy\in\wl_m(m)\enskip \forall t\enskip
x_t<t<y_t\rightarrow  \W(m,m,0,0,0,t).
\end{multline*}
\end{itemize}

Our last two axioms will ensure that the worldlines of accelerated
observers are ``tame" enough, e.g., they have velocities at each
moment. In \SpecRel, the worldview transformations between inertial
observers are affine functions, the next axiom will state that the
worldview transformations between accelerated observers are
approximately affine, wherever they are defined.

\begin{itemize}
\item[\underline{\AxDif:}] The worldview transformations have
affine approximations at each point of their domain (i.e., they are
differentiable):
\begin{equation*}\forall m,k\in\Ob\enskip\forall \vx\in\Dom(\w_{mk})\enskip\exists
\mbox{ affine } A \enskip \w_{mk}\sim_{\vx} A,\footnote{The quantifier
  ``$\exists \text{ affine } A$'' looks like a second-order logic one,
  but truly it is a first-order logic quantifier because every affine
  map from $\Q^4$ to $\Q^4$ can be represented by a $4\times4$ matrix,
  i.e., $16$ elements of $\Q$, together with an $\vx\in\Q^4$.}
\end{equation*}
\end{itemize}
where $\Dom (R)$, the domain of a binary relation $R$, is defined
as:
\begin{equation*}
 \Dom (R) \de \{\, x \::\: \exists y\enskip \langle
x,y\rangle \in R \,\}.
\end{equation*}

We note that \AxDif\ implies that the worldview transformations are
functions with open domains. However, if the numberline has gaps,
still there can be crazy motions. Our last assumption is an axiom
scheme supplementing \ax{AxDf} by excluding these gaps.

\begin{itemize}\label{p-cont}
\item[\underline{\AxCont:}] Every subset of \Q\ which is definable, bounded and nonempty
   has a supremum.
\end{itemize}

\noindent In \ax{Cont} ``definable" means ``definable in the language
of \AccRel, parametrically.'' For a precise formulation of \ax{Cont},
see \cite[p.692]{MNSzFP} or \cite[\S 10.1]{SzDis}.

That \ax{Cont} requires the existence of supremum only for sets
definable in the language of \ax{AccRel} instead of every set is
important not only because by this trick we can keep our theory within
FOL (which is crucial in our foundational setting), but also because it
makes this postulate closer to the physical/empirical level. This
is true because \ax{Cont} does not speak about ``any fancy subset'' of
the quantities, but just about those ``physically meaningful'' sets which can be
defined in the language of our (physical) theory.

Our axiom scheme of continuity (\ax{Cont}) is a ``mathematical
axiom" in spirit. It is Tarski's first-order logic version of
Hilbert's continuity axiom in his axiomatization of geometry, see
\cite[pp.161-162]{Gol}, fitted to the language of \ax{AccRel}.

When $\Q$ is the usual real number-line, \ax{Cont} is automatically
true.

Adding this five axioms to \ax{SpecRel}, we get the following axiom
system for accelerated observers:
\begin{equation*}
\AccRel\de\SpecRel\cup\{\AxCmv, \AxEv^-, \AxSelf^-,
\AxDif\}\cup\AxCont.
\end{equation*}

\noindent The explicit introduction and development of \AccRel\ as a
theory in its own right is a contribution of our group. As an
example we show that the so-called \textit{twin paradox} can be
naturally formulated and proved in $\AccRel$.  More importantly, the
details of the twin paradox (e.g., who sees what, when) can be
analyzed with the clarity of logic, see \cite[pp.139--150]{pezsgo} for
part of such an analysis.

According to the twin paradox, if a twin makes a journey into space
(accelerates), he will return to find that he has aged less than his
twin brother who stayed at home (did not accelerate). We formulate
the twin paradox in our FOL language as follows.

\begin{itemize}
\item[\underline{\ax{TwP}:}]  Every inertial observer $m$
 measures at least as much time as any other observer $k$ between any
 two events $e_1$ and $e_2$ in which they meet; and they measure the same time iff they
 have encountered the very same events between $e_1$ and $e_2$:
\begin{multline*}
\forall m \in \IOb\enskip \forall k\in \Ob\enskip \forall
\vx\vx'\vy\vy' \enskip x_t<y_t \land x'_t<y'_t \,\land \\
m,k\in\ev_m(\vx)=\ev_k(\vx')\land m,k\in\ev_m(\vy)=\ev_k(\vy') \then
y'_t-x'_t\le y_t-x_t \\ \land\big( y'_t-x'_t=y_t-x_t \iff
enc_m(\vx,\vy)=enc_k(\vy',\vy') \big),
\end{multline*}
\end{itemize}
where  $enc_m(\vx,\vy)=\{\ev_m(\vz): \W(m,m,\vz) \land x_t\le z_t\le y_t\}$.

\begin{thm}\label{thm-twp}
${}$
\begin{itemize}
\item[(i)]
$\AccRel\models\ax{TwP}$.
\item[] {\rm Moreover, $\AccRel-\ax{AxDf}\models\ax{TwP}$.}
\item[(ii)] $\AccRel-\ax{Cont}\not\models\ax{TwP}$.
\item[] {\rm Moreover, for any Euclidean ordered field $\F$ different
  from the field of real numbers there is a model $\M$ such that its
  field reduct $\langle \Q,+, \cdot\rangle$ is $\F$ and
  $\M\models(\AccRel - \AxCont)$, but $\M\not\models\ax{TwP}$.}
\end{itemize}
\end{thm}

For the proof of Theorem~\ref{thm-twp}, see \cite{MNSzFP} or
\cite[\S 7]{SzDis}.

The second part of Theorem \ref{thm-twp} implies that \ax{Cont}
cannot be replaced with the whole FOL theory of real numbers in
\ax{AccRel} if we do not want to lose \ax{TwP} from its
consequences.

\input epsf
\begin{figure}[hbt]
\small
\setlength{\unitlength}{0.8 truemm}
\begin{center}
\begin{picture}(150,80)(0,0)
\put(15,75){\makebox(0,0){$m$}}
\put(7,25){\makebox(0,0){$\vx$}}
\put(7,67){\makebox(0,0){$\vy$}}
\put(30,48){\makebox(0,0){$k$}}
\put(91,33){\makebox(0,0){$\vx'$}}
\put(91,57){\makebox(0,0){$\vy'$}}
\put(90,75){\makebox(0,0){$k$}}
\put(75,46){\makebox(0,0){$m$}}
\put(73,5){\makebox(0,0)[r]{$m$: ``non-moving'' (inertial) brother}}
\put(82,5){\makebox(0,0)[l]{$k$: traveling  (accelerated) brother}}
\epsfysize = 80\unitlength
\epsfbox{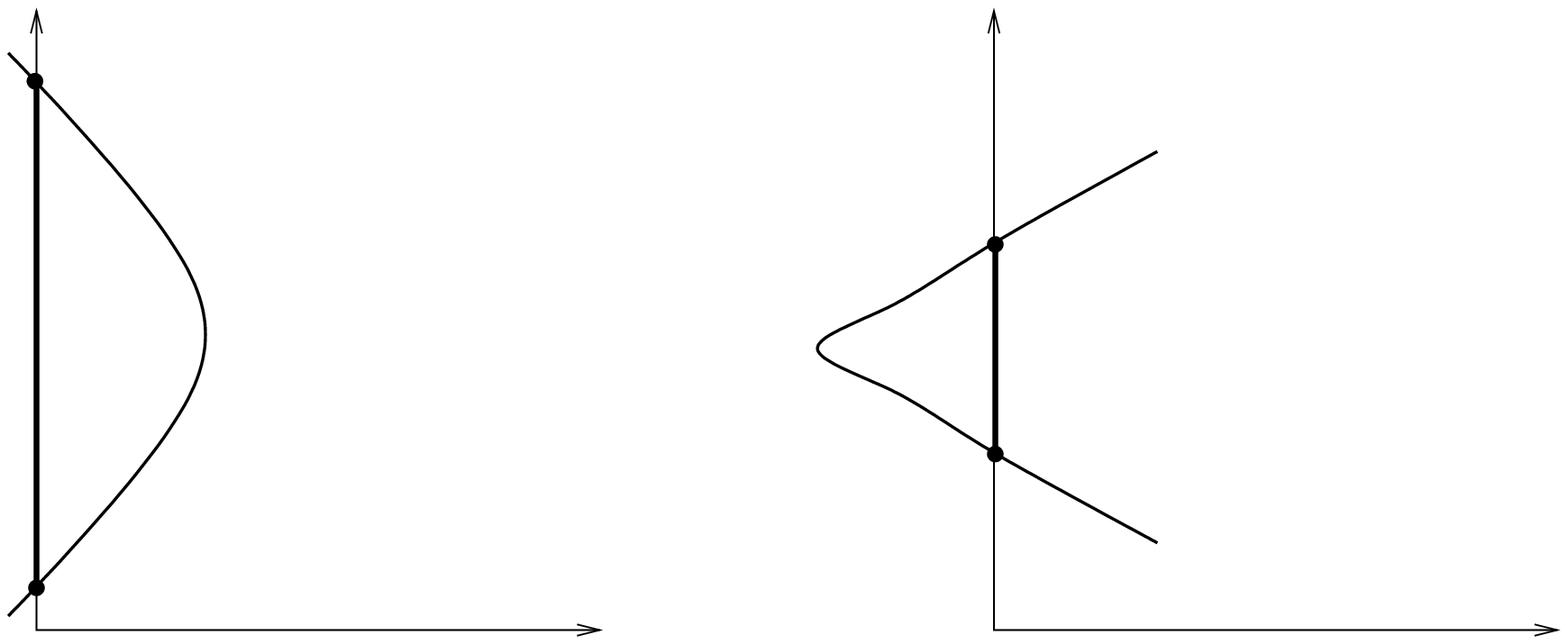}
\end{picture}
\end{center}
\caption{\label{twp-f} The ``twin paradox''}
\end{figure}

Adding the axiom schema \AxCont\ to our axioms in \AccRel\
represents a first step in the direction pursued in the so-called
nonstandard-time logics of time approach represented by
\cite{AGMNS95}, \cite{Sa86}.

 All this enables us to ``import'' just as much of any field of
 mathematics, e.g., mathematical analysis into our first-order theory
 \AccRel\ of accelerated observers as we need. Explicit and detailed
 elaborations of these ideas to situations similar to our present one
 (theory of accelerated observers) can be found in the above quoted
 \cite{AGMNS95}, \cite{Sa86} and in the works quoted therein. For
 developing \AccRel\ further, it is a distinct possibility to adopt
 the methods of the works just quoted to the framework of
 \AccRel\ (extended with first-order logic sorts to treat second-order
 logic objects, just like in Henkin-style second-order logic).

\section{An axiomatization of general relativity in FOL}

The theory of accelerated observers \AccRel\ speaks about two kinds
of observers, inertial ones and accelerated ones. Some of the axioms
are postulated for inertial observers only (such is, e.g., \AxSimd),
some of the axioms  apply to all observers (such is, e.g.,
$\AxSelf^-$), and there is one axiom, \AxCmv, which talks about both
of them. We get the axiom system \GenRel\ for general relativity by
stating the axioms of \AccRel\ in a generalized form in which they
are postulated for all observers, inertial and accelerated ones
equally. In other words, we will change all axioms of \AccRel\ in
the same spirit as $\AxSelf^-$ and $\AxEv^-$ were obtained from
\AxSelf\ and \ax{AxEv}, respectively.
This kind of change \ax{AccRel} $\mapsto$ \ax{GenRel} can be
regarded as a ``democratic revolution'' with the slogan ``all
observers are equivalent, the same laws should apply to all of
them.'' Here ``law'' translates as ``axiom.'' This idea originates with
Einstein (cf.\ his book \cite[Part II, ch.18]{Ein}). In
\cite[pp.58(ch.18),88(ch.28)]{Ein}, Einstein calls this idea of
``democratic revolution of observers'' the ``General Principle of
Relativity.'' Below, we implement Einstein's idea in logic
(particularly in FOL).

For simplicity, we will use an equivalent version of the symmetry
axiom \AxSimd\ (see \cite[Thm.2.8.17(ii), p.138]{pezsgo} or
\cite[Thm.3.1.4, p.21]{SzDis}), and we will require the speed of photons
to be 1 in $\AxPh^-$ (as opposed to requiring it in $\AxSimd^-$).

We will need the notion of velocity for curved worldlines. This
notion will be based on the usual notion of the affine
approximation.  First we define the affine approximation (also
called linear approximation or differential) of $f$ at $\vx$,
denoted by $\Apr(f,\vx)$.
\begin{equation*}
\Apr(f,\vx)=g\defiff f\sim_{\vx} g\mbox{ and } g\mbox{ is affine}.
\end{equation*}
Let $f:\Q\to\Q^n$ be an affine map. We will use the following
auxiliary notation for the velocity of $f$:
\begin{equation*}
v(f)\de\left\{
\begin{array}{cl} f(1)-f(0) & \text{ if } n=1,\\
\frac{f(1)_s-f(0)_s}{f(1)_t-f(0)_t} & \text{ if } n>1 \text{ and }
f(1)_t\ne f(0)_t,\\ \text{undefined} & \text{ otherwise}.
\end{array}
\right.
\end{equation*}
To define the velocity of body $b$ according to observer $m$, first
we define the time-parameterized worldline of $b$ (parameterized by
the time of $m$):
\begin{multline*}
\wl_m(b)(t)=\vx \defiff \\\W(m,b,\vx)\land x_t=t \land \big(\forall\vy\enskip
\W(m,b,\vy)\land y_t=t\then \vx=\vy\big).
\end{multline*}
By this definition, the time-parameterized worldlines are partial
functions from $\Q$ to $\Q^4$. Now we can define the velocity of
body $b$ according to observer $m$ as follows:
\begin{equation*}
v_m(b,\vx)\de v(\Apr(\wl_m(b),\vx)).
\end{equation*}

The behavior of observer $k$'s clock as seen by observer $m$ is defined
as follows:
\begin{multline*}
\cl_m(k)(t)=t' \defiff \exists \vx\vx'\enskip \W(m,k,\bar x)\land\ev_m(\vx)=\ev_k(\vx')\land
\\ t=x_t\land t'=x'_t\land\big(\forall\vz\enskip \W(m,k,\vz)\land
z_t=t\then \vx=\vz\big).
\end{multline*}
By this definition, $\cl_m(k)$ is a function relating $t'$ to $t$ if
$k$'s clock shows $t'$ when $m$'s clock shows $t$. That is,
$\cl_m(k)$ is the time $k$'s clock shows at $t$ according to $m$'s
clock. Thus, e.g., $v(\Apr(\cl_m(k),t))=2$ means that at $t$
(according to $m$'s clock) $k$'s clock runs twice as fast as $m$'s.
Now we are ready to state our axioms for general relativity.

\begin{itemize}
\item[\underline{$\AxPh^-$:}] The velocity of photons an observer ``meets" is 1
when they meet, and it is possible to send out a photon in each
direction where the observer stands, i.e.,
\begin{multline*}\forall k\enskip\forall p\in\Ph\enskip
\forall\vx\in\wl_k(k)\cap\wl_k(p)\enskip |v_k(p,\bar x)|=1 \text{\quad
  \rm and }\\ \forall k\enskip\forall\bar x\in\wl_k(k)\enskip\forall
v\in\Q^3\enskip |v|=1\rightarrow\exists p\in\Ph\cap\ev_k(\vx)\enskip
v_k(p,\vx)=v.
\end{multline*}
\item[\underline{$\AxSimd^-$:}] Meeting observers see each other's clocks slow
down with the same rate, i.e.,
\begin{multline*}\forall mk\vx\vy\quad
m,k\in\ev_m(\vx)=\ev_k(\vy)\\\then\enskip
v(\Apr\big(\cl_m(k),x_t)\big)=v\big(\Apr(\cl_k(m),y_t)\big).
\end{multline*}
\end{itemize}

We introduce an axiom system for general relativity as the collection
of the following axioms:
\begin{equation*}
\GenRel\de\{\AxField,\AxPh^-,\AxEv^-,\AxSelf^-,\AxSimd^-,\AxDif\}\cup\AxCont.
\end{equation*}

Axiom system \ax{GenRel} contains basically the same axioms as
\ax{SpecRel}, the difference is that they are assumed only locally
but for all the observers.  Axiom \ax{AxDf} also fits into this
picture. By Theorem \ref{thm-poi}, the worldview transformations
between inertial observers are affine ones, and \ax{AxDf} is the
localization of this statement assumed for all the observers.
\ax{Cont} is a property of real numbers that we could have assumed
in \ax{SpecRel} but we did not assume it because it was not needed.

The following theorem states that our axiom system \ax{GenRel}
captures general relativity in that its models are exactly the
spacetimes of usual general relativity. For the notion of a
Lorentzian manifold we refer to \cite[p.55]{dinverno},
\cite[p.241]{MTW} and \cite[sec.3.2]{logst}.

\begin{thm}[Completeness theorem]\label{grcomp-thm}
$\GenRel $ is complete with respect to its standard models, i.e., to
  Lorentzian Manifolds over real closed fields.
\end{thm}

This theorem can be regarded as a completeness theorem in the
following sense. Let us consider Lorentzian manifolds as intended
models of \ax{GenRel}. How can we do that? We give a method for
constructing a model of \ax{GenRel} from each Lorentzian manifold;
and conversely, we show that each model of \ax{GenRel} is obtained
in this way from a Lorentzian manifold.  By the above, we defined what
we mean by a formula $\varphi$ in the language of \ax{GenRel} being
valid in a Lorentzian manifold, or in all Lorentzian manifolds. Then
completeness means that for any formula $\varphi$ in the language of
\ax{GenRel}, we have $\ax{GenRel}\vdash \varphi$ iff $\varphi$ is
valid in all Lorentzian manifolds over real closed fields.  That is
completely analogous to how Minkowskian spacetimes were
regarded as intended models of \ax{SpecRel} in the completeness
theorem of \ax{SpecRel}, see \cite[Thm.11.28, p.681]{logst} and
\cite[\S 4]{Mphd}. For more on the proofidea for our completeness
theorem for \GenRel, cf.\ also \cite[items 11.28-11.30,
pp.681-2]{logst}.

Our theory \ax{GenRel} was obtained from \ax{AccRel} by getting rid
of the concept of inertiality on the level of our axioms. However, we
can recover this concept. We call the worldline of observer $m$
\textit{timelike geodesic}, if each of its points has a neighborhood
within which this observer ``maximizes measured time" between any
two encountered events, i.e.,
\begin{multline*}
   \forall\vz \in \wl_m(m)\;\exists \delta \enskip \delta>0\, \land \\
   \forall k\vx\vy \enskip |\vx-\vz|<\delta\land |\vy-\vz|< \delta \land\Ob(k)\land \vx,\vy\in
   \wl_m(m)\cap\wl_m(k)\\\land
  \big(\forall \vw\in \wl_m(k) \enskip |\vw-\vz|<\delta \big) \then |x_t-y_t| \geq
   \left|\w_{mk}(\vx)_t-\w_{mk}(\vy)_t\right|.
\end{multline*}
In this case we also say that observer $m$ is an {\it inertial} body
(but not necessarily an inertial observer).  This definition is
justified by the twin paradox theorem of \ax{AccRel}, see
Theorem~\ref{thm-twp}. This theorem says that in the models of
\ax{AccRel} the worldlines of {\it inertial} observers are timelike
geodesics in the above sense.

According to the definition above, if there are only a few observers,
then it is not a big deal that the worldline of $m$ is a time-like
geodesic (it is easy to be maximal if there are only a few to be
compared to). To generate a real competition for the rank of having
a timelike geodesic worldline, we postulate the existence of many
observers by the following axiom scheme of comprehension.

\begin{itemize}
\item[\underline{\ax{Compr}:}] For any parametrically definable
  timelike curve in any observer's worldview, there is another observer
  whose worldline is the range of this curve.
\end{itemize}
A precise formulation of \ax{Compr} can be obtained from
that of its analogue in \cite[p.679]{logst}.

The assumption of axiom schema \ax{Compr} guarantees that our
definition of geodesic coincides with that of the literature on
Lorentzian manifolds. Therefore we also introduce the following theory:
\begin{equation*}
\ax{GenRel^+}\de \ax{GenRel}\cup\ax{Compr}.
\end{equation*}
So in our theory \ax{GenRel^+}, our notion of timelike geodesic
coincides with its standard notion in the literature on general
relativity.  All the other key notions of general relativity, such as
curvature or Riemannian tensor field, are definable from timelike
geodesics. Therefore we can treat all these notions (including the
notion of metric tensor field) in our theory \ax{GenRel^+} in a
natural way.

In general relativity Einstein's equations give the connection
between the geometry of the spacetime and the energy-matter
distribution (given by the energy-momentum tensor field). Since in
\ax{GenRel^+} all the geometric notions of the spacetime are
definable, we can use Einstein's equation as a definition of the
energy-momentum tensor, see, e.g., \cite{Benda} or \cite[\S 13.1,
  p.169]{dinverno}, or we can extend the language of \ax{GenRel^+} by
the concept of energy-momentum tensor and assume Einsten's equations
as axioms. As far as we do not assume anything more from the
energy-momentum tensor than its connection to the geometry described
by Einstein's equations, there is no real difference in these two
approaches. In both approaches we can add any extra condition about
the energy-momentum tensor to our theory, e.g., the dominant energy
condition or that the spacetimes are vacuum solutions.

\bigskip\bigskip
\noindent {\sc H.~Andr\'eka, J.~X.~Madar{\'asz}, \\
I. N\'emeti, G. Sz{\'e}kely\\} Alfr{\'e}d R{\'e}nyi
Institute of Mathematics\\ of the Hungarian Academy of Sciences\\
Budapest P.O.
Box 127, H-1364 Hungary\\ {\tt andreka@renyi.hu, madarasz@renyi.hu\\
nemeti@renyi.hu,
  turms@renyi.hu}


\begin{thebibliography}{}

\bibitem{AGMNS95} Andr{\'e}ka, H., V.~Goranko, Sz.~Mikul{\'a}s,
  I.~N{\'e}meti, and I.~Sain. \newblock {Effective first order
    temporal logics}. \newblock pp.~51--129. In {L}.~{B}olc and
  {A}.~{S}za{\l}as (eds.)  {\em Time and Logic, a computational
  approach}
  {UCL}, {L}ondon, 1995.

\bibitem{pezsgo} Andr{\'e}ka, H., J.~X.~Madar{\'a}sz, and
  I.~N{\'e}meti. \newblock {\it On the logical structure of
    relativity theories}. \newblock E-book, {A}lfr{\'e}d {R}{\'e}nyi
  {I}nstitute of {M}athematics, {B}udapest, 2002.  With contributions from
  {A}.~{A}ndai, {G}.~{S}{\'a}gi, {I}.~{S}ain, and {C}s.~{T}{\H o}ke. \newblock
  http://www.math-inst.hu/pub/algebraic-logic/olsort.html. 1312 pp.

\bibitem{logst} Andr{\'e}ka H., J.~X. Madar{\'a}sz, and I.~N{\'e}meti.
  \newblock Logic of space-time and relativity theory.  \newblock In
  {M}.~{A}iello, {I}.~{P}ratt-{H}artmann, and {J}.~van {B}enthem, (eds.)
  \emph{Handbook of spatial logics}, pp.~607--711. {S}pringer-Verlag,
  {D}ordrecht, 2007.

\bibitem{AMNsamples}
Andr{\'e}ka H., J.~X. Madar{\'a}sz, and I.~N{\'e}meti.
\newblock Logical axiomatizations of space-time. {S}amples from the literature.
\newblock In {A}.~{P}r{\'e}kopa and {E}.~{M}oln{\'a}r (eds.) {\em Non-{E}uclidean
  geometries}, pp.~155--185. {S}pringer-Verlag, {N}ew {Y}ork, 2006.

\bibitem{WKU09} Andr{\'e}ka, H., J. X.~Madar{\'a}sz, I.~N{\'e}meti,
  P.~ N{\'e}meti,  and  G.~Sz{\'e}kely. {Vienna Circle and logical
    analysis of relativity theory}. \newblock In {F}.~{S}tadler (ed.)
  {\em {W}iener {K}reis und {U}ngarn}. \newblock
  {V}er{\"o}ffentlishungen des {I}nstituts {W}iener {K}reis,
  {S}pringer--Verlag, to appear.

\bibitem{AMNSzKalmar} Andr\'eka, H., J.~X. Madar\'asz, I.~N\'emeti, and
  G.~Sz\'ekely.  \newblock A logical investigation of inertial and
  accelerated observers in flat space-time.  \newblock In
  {F}.~{G}{\'e}cseg, {J}.~{C}sirik, and {G}y. {T}ur{\'a}n  (eds.) {\em {K}alm\'ar
    Workshop on Logic and Computer Science}, pp.~45--57, {JATE}
  {U}niversity of {S}zeged, {S}zeged, 2003.

\bibitem{AMNSzCN} Andr\'eka, H., J. X. Madar\'asz, I. N\'emeti, and
  G. Sz\'ekely. \newblock {\em Relativity Theory on Logical
    Grounds}. \newblock Course Notes, {B}udapest 2010. \newblock
  http://www.math-inst.hu/pub/algebraic-logic/kurzus10/kurzus10.htm

\bibitem{Benda} Benda, T.  \newblock A formal construction of the
  spacetime manifold.  \newblock {\em J. Phil. Logic}, 37(5):441--478,
  2008.

\bibitem{CK}  Chang, C.~C., and H.~J.~Keisler.  \newblock {\em Model
  theory}.  \newblock {N}orth-{H}olland {P}ublishing Co., {A}msterdam, 1990.

\bibitem{dinverno} d'Inverno, R. \newblock {\em Introducing
  {E}instein's relativity}. \newblock {O}xford {U}niversity Press,
  {N}ew {Y}ork, 1992.

\bibitem{Ein} Einstein, A. \newblock {\em Relativity. The Special and the General Theory.}
\newblock Penguin Classics, 2006. Translated by W. Lawson (original publication in German 1921).

\bibitem{End} Enderton, H.~B. \newblock {\em A mathematical
  introduction to logic}. \newblock {A}cademic Press, {N}ew {Y}ork,
  1972.

\bibitem{Gol} Goldblatt, R. \newblock {\em Orthogonality and spacetime
geometry.} \newblock Springer-Verlag, New York, 1984.

\bibitem{HMT} Henkin, L., J.~D.~Monk, and A.~Tarski. \newblock {\em
  Cylindric Algebras. Part I}. {N}orth-{H}olland {P}ublishing Co.,
  {A}msterdam, 1971.

\bibitem{Hodges} Hodges, W. \newblock {\em Model Theory}, Encyclopedia
  of Mathematics and its Applications, 42. \newblock Cambridge
  Univ.\ Press, Cambridge, 1993.

\bibitem{Mphd}  Madar{\'a}sz, J.~X.  \newblock {\em Logic and
  Relativity (in the light of definability theory)}.  \newblock PhD
  thesis, {E}{\"o}tv{\"o}s {L}or{\'a}nd Univ., {B}udapest, 2002.
  \newblock http://www.math-inst.hu/pub/algebraic-logic/Contents.html.

\bibitem{MNSzFP} Madar{\'a}sz, J.~X., I.~N{\'e}meti, and
  G.~Sz{\'e}kely.  \newblock Twin paradox and the logical foundation
  of relativity theory.  \newblock {\em Found. Phys.}, 36(5):681--714,
  2006.

\bibitem{MNSzTamara} Madar{\'a}sz, J.~X., I.~N{\'e}meti, and
  G.~Sz{\'e}kely.  \newblock First-order logic foundation of
  relativity theories.  \newblock In {D}.~{G}abbay et~al. (eds.) {\em
    Mathematical problems from applied logic II., New Logics for the
    XXI-st Century}, pp.~217--252. Springer-Verlag, New York, 2007.

\bibitem{MTW} Misner, C.~W., K.~S.~Thorne, and J.~A.~Wheeler.
  \newblock {\em Gravitation}.  \newblock {W}. {H}. {F}reeman and Co., {S}an
  {F}rancisco, 1973.

\bibitem{Rosinger08} Rosinger E.~E. \newblock {\em Two Essays on the Archimedean
  versus Non-Archimedean Debate}. \newblock  2008, arXiv:0809.4509v3.

\bibitem{Rosinger09} Rosinger E.~E. \newblock {\em Special Relativity
  in Reduced Power Algebras}. \newblock 2009, arXiv:0903.0296v1.

\bibitem{Sa86} Sain, I. \newblock {\em Dynamic logic with nonstandard
  model theory}. \newblock Dissertation, 1986.

\bibitem{SzDis} Sz\'ekely, G. \newblock {\em First-Order Logic
  Investigation of Relativity Theory with an Emphasis on Accelerated
  Observers}. \newblock PhD thesis, {E}{\"o}tv{\"o}s {L}or{\'a}nd
  Univ., {B}udapest, 2009. \newblock
  http://www.renyi.hu/~turms/phd.pdf.

\bibitem{Takeuti} Takeuti, G. \newblock \emph{Proof Theory} \newblock
  {N}orth-{H}olland {P}ublishing Co., {A}msterdam, 1987.

\bibitem{TarskiElge} Tarski, A. \newblock What is elementary geometry?
  In {L}.~{H}enkin, {P}.~{S}uppes, and {A}.~{T}arski (eds.) {\em The
    axiomatic method. {W}ith special reference to geometry and
    physics.} \newblock pp.~16--29, North-Holland Publishing Co.,
  Amsterdam, 1959.


\end{thebibliography}
\end{document}